\documentclass[conference]{IEEEtran}
\IEEEoverridecommandlockouts
\usepackage{cite}
\UseRawInputEncoding  % Hao added this line to get rid of "inputenc error invalid UTF-8 for bib.tex
\usepackage{amsmath,amssymb,amsfonts}
\usepackage{algorithmic}
\usepackage[hyphens]{url} % Hao Luan added this line to address url doesn't wrap in ref
\usepackage{graphicx}
\usepackage{textcomp}
\usepackage{xcolor}
\def\BibTeX{{\rm B\kern-.05em{\sc i\kern-.025em b}\kern-.08em
    T\kern-.1667em\lower.7ex\hbox{E}\kern-.125emX}}
\begin{document}

\title{Combinatorics and Geometry for the Many-ported, Distributed and Shared Memory Architecture }
\author{Hao Luan, \emph{Member, IEEE,} Alan Gatherer, \emph{Fellow, IEEE} \\
	Futurewei Technologies Inc. Plano, Texas, USA 75024\\
	\{hao.luan, alan.gatherer\}@futurewei.com}

\maketitle
\begin{abstract}
Manycore SoC architectures based on on-chip shared memory are preferred for flexible and programmable solutions in many application domains. However, the development of many ported memory is becoming increasingly challenging as we approach the end of Moore’s Law while systems requirements demand larger shared memory and more access ports. Memory can no longer be designed simply to minimize single transaction access time, but must take into account the functionality on the SoC. In this paper we examine a common large memory usage in SoC, where the memory is used as storage for large buffers that are then moved for time scheduled processing. We merge two aspects of many ported memory design, combinatorial analysis of interconnect, and geometric analysis of critical paths, extending both to show that in this case the SoC performance benefits significantly from a hierarchical, distributed and staged architecture with lower-radix switches and fractal randomization of memory bank addressing, along with judicious and geometry aware application of speed up. The results presented show the new architecture supports 20\% higher throughput with 20\% lower latency and 30\% less interconnection area at approximately the same power consumption. We demonstrate the flexibility and scalability of this architecture on silicon from a physical design perspective by taking the design through layout. The architecture enables a much easier implementation flow that works well with physically irregular port access and memory dominant layout, which is a common issue in real designs.
\end{abstract}


\section{Introduction}\label{sec-Intro}
Many core SoC architectures based on many ported on-chip shared memory are the preferred
backbone for flexible and programmable solutions in many computationally intensive application domains, including machine learning \cite{10.1145/2996864}, high performance wireless infrastructure \cite{Marvell/Octeon_CNF95xx}, and embedded processing \cite{6800311}. For the wireless application on the order of 30 to 100 processing blocks on a single chip are not uncommon to fulfill the overall processing needs. The  Marvel CNF73xx LTE baseband SoC in \cite{Marvell/Octeon_CNF73xx} has similar architecture choices with a common on chip memory of 8 Mega bytes shared among over 30 heterogeneous accelerators and DSP cores.  With the introduction of 5G/New Radio, the size of the shared memory has been tripled to 24 Mega byte and shared among 42 DSP cores and some other base band hardware accelerators \cite{Marvell/Octeon_CNF95xx}.\\
Despite the requirement of increasingly large shared memory the performance of the SoC will suffer if the memory throughput to the compute elements (CEs) in the SoC decreases. Many CEs, especially if they have no software optimized prefetch, will suffer rapid performance loss if the average latency to memory increases. Average memory latency is the critical metric because CEs are scheduled to run functions that may take thousands of clock cycles to complete and similarly consume/produce thousands of Bytes of data. It is only the completion time of the function, and therefore the average memory access latency, that matters to the performance of the CE. Finally an important practical metric is the effort expended in re-layout when the memory size changes, the memory bounding box shape alters or the number of ports changes. A good memory design will cope easily with such changes at layout. \\
In this paper we address all of the above concerns using a combination of combinatorial analysis of interconnect, and geometric analysis of critical paths, and show that the SoC performance benefits significantly from a hierarchical, distributed and staged architecture with lower-radix switches and fractal randomization of memory bank addressing, along with judicious and geometry aware application of speed up. The results presented show the new architecture supports 20\% higher throughput with 20\% lower latency and 30\% less interconnection area at approximately the same power consumption. \\
The rest of the paper is organized as follows: Section \ref{sec-Related} surveys
related work. Section \ref{sec-Arch} presents our architecture,
describing in detail  its key technologies. In Section \ref{sec-Results}, we share relevant results to validate the effectiveness of our approach, while Section \ref{sec-Conclusion} summarizes our conclusions and findings.

\section{Related Work}\label{sec-Related}
Shared memory architecture has been explored in multi-core and many core settings inside a cluster \cite{5763085} using Mesh-of-Trees (MoT) topology giving single cycle access but at a very limiting frequency of 310 MHz and 256 Kbytes shared L1 memory in 65nm. A 4x8 MoT is also used in \cite{5763085} where every level of fan-out dilutes the traffic conflicts by half but doubles the interconnect. As analyzed in \cite{4019494}, even though it can achieve 90\% throughput the number of wires increases $O(4n^2)$ and wire crossings  increase $O(16n^4)$ where $n$ doubles every routing stage. \\
In \cite{6800311} the computing cluster has a 2 Mbytes shared memory in which wire density is reduced by dividing the memory into two sides and introducing a 4-by-2 routing switch for compute element pairs. Each pair has two memory buses (one for each side), which can be utilized in parallel by the two CEs. The 4-by-2 routing switch reduces the interconnects from around 64 million connections to around 16 million, a 4X reduction, but the ideal throughput drop per port is only $6.25\%$. This enables 2MB of 550MHz shared memory in 28 nm HP. Some of this performance was due to the high regularity of layout where all the PEs are sandwiched between the memory banks located on the both sides. This regularity of the geometry can be very hard to find once we have many master ports and sea of memory banks spread all over the silicon. \\
In Dadiannao\cite{10.1145/2996864}, a Fat tree topology is used inside each tile and the wires occupy almost half of the area. It can be utilized as a hierarchical architecture but global resource sharing is limited due to its intrinsic NUMA nature, which is not suitable for a shared memory architecture.\\
Flattened butterfly \cite{4408254} topology can be a good candidate for the shared memory architecture if the size of shared memory is not too big and geometry can be regular, because it collapses all vertical switches and connections but leaves only horizontal connections among collapsed switches.  However there are  physical challenges once the size of shared memory grows bigger or the physical layout becomes irregular, which is driven by the surrounding heterogeneous PEs. \\
From the above examples we see that a judicious reduction in connectivity at the right places in the many port interconnect can have a dramatic improvement on area and performance with only little degradation in latency and in this paper we take steps to quantify and understand how this may best be done. To introduce the relevant concepts Fig. {\ref{shared-flat-arch}} gives an example of \textbf{staged speed-up} where $n$ master ports sharing a logical memory connect to $k$ memory ports and each memory port can connect $r$ memory banks, which we define as memory speed-up of $r$ because transactions that would normally collide at a memory port may now be able to continue onto different banks. When $r = 1$, this interconnect network is reduced to what we will call a \textbf{conventional} $n \times k$ full-crossbar topology. Fig. {\ref{Convention-Shared-layout} shows a conventional topology layout with 32 masters and 8 mega bytes of memory. We can observe that non uniform latency to the shared memory architecture is mainly caused by the big physical footprint of many instances of memory. Non-uniform access delay to area A, B and C is unavoidable and will worsen as the geometry becomes more irregular. Therefore, we must address this at the architecture level.\\
	\begin{figure}
		\includegraphics[width=0.8\linewidth, height=5cm]{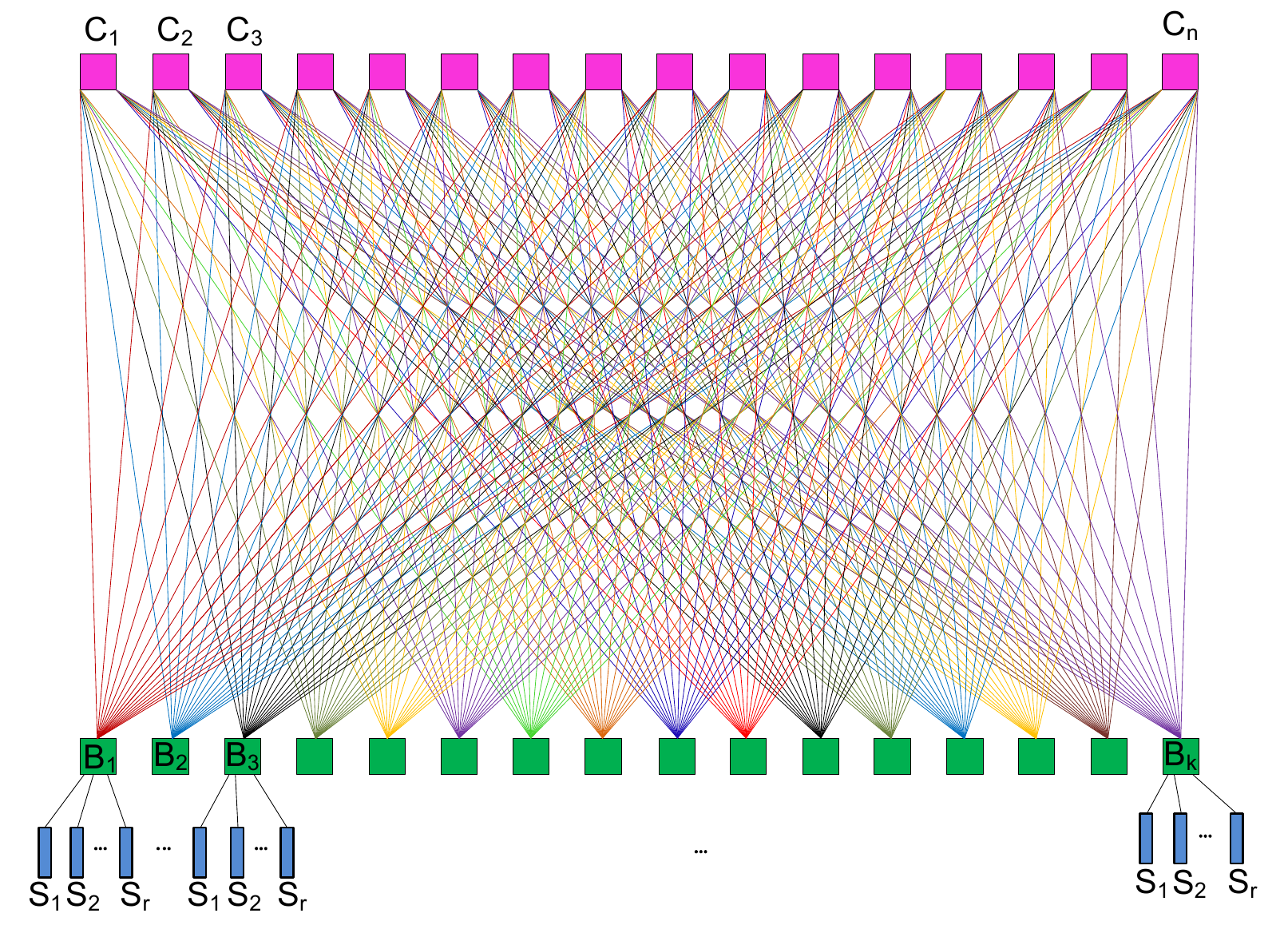}
		\centering
		\caption{A Flat Full Cross Bar Interconnect Network}
		\label{shared-flat-arch} 
	\end{figure}
	\begin{figure}
		\includegraphics[width=0.6\linewidth,height=0.5\linewidth]{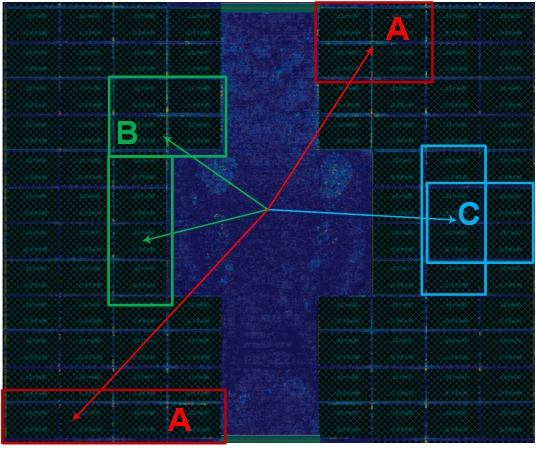}
		\centering
		\caption{An Example of Shared Memory Layout with Low Silicon Utilization}
		\label{Convention-Shared-layout} 
	\end{figure}
	Interconnect has replaced transistors as the main determinant of performance, and this "tyranny of interconnection" will only escalate \cite{1166548} as interconnect resistance and capacitance (RC) increasingly dominate delay \cite{Radamson_Zhang_2017}. As pointed out in \cite{4568311}, the number of wire crossings is also critical for the effectiveness and performance of a specific topology. In Fig. \ref{shared-flat-arch} if we assume $n=k$ there are $O(n^2)$ wires between master ports and memory ports, and there are $O(n^4)$ wire crossings as analyzed in Formula (\ref{equ-radix-N-fully-connected}). The number of wire crossings leads to routing congestion and low silicon utilization as shown as the "swimming pool" area in the middle of Fig. \ref{Convention-Shared-layout}, where the utilization of interconnect is only a single digit.\\

\section{Architecture}\label{sec-Arch}
Our architecture takes the concepts of distributed shared memory \cite{494605} and extends it with the key technologies described below. All the technologies are effective individually, but provide the best benefits and results when they are applied together. For reasons that will become clear we call this memory architecture the Distributed Shared Memory Controller {\textendash} DSMC \cite{luan_gatherer_DSMC_2020}. \\
\subsection{Mathematical Modeling and Analysis on Speed-ups}\label{Math_Analysis}
\color{black}
For the topology shown in Fig.\ref{shared-flat-arch}, there are $m=k\cdot r$ memory banks in total. Assume all masters and memory banks are running at the same frequency and all of the master requests are statistically independent and identical with probability $P_a$.  Then the probability a specific master port accesses a specific slave port is  {\em $\frac{P_a}{k}$}. Define $P\{q\}$ to be the probability that there will be {\em q} requests at a specific slave port\cite{7167206}. 
\begin{equation}
\mathbf P\{q\}=\begin{pmatrix} n\\q\end{pmatrix}\bigg(\frac{P_a}{k}\bigg)^q\bigg(1-\frac{P_a}{k}\bigg)^{n-q} 
\end{equation}
The expected number of requests that will leave the slave port towards the memory banks depends on how the slave port processes requests. If there are $q<r$ requests to a slave port then the utilization of the port will be $f_r(q)$ due to random contentions at memory banks
\begin{equation}
f_r(q)= r\Bigl(1-\Bigl(\frac{r-1}{r}\Bigr)^q\Bigr) \label{equ-expected-speedup-network}
\end{equation}
When there are $q \geq r$ requests to a slave port we assume that $r$ requests are randomly chosen and kept for processing while the others are back pressured. This means that the utilization is the same as for $r$ requests, $f_r(r)$. So the expected utilization of the slave port is:
\begin{align}
\mathbf E(k,n,r) &=\sum_{q=0}^{r-1}f_r(q)P\{q\} + \sum_{q=r}^{n}f_r(r)P\{q\} \\
&=\sum_{q=0}^{r-1}f_r(q)P\{q\} +f_r(r)(1-\sum_{q=0}^{r-1}P\{q\}) \\ 
&=r\Bigg[\Bigl(1-\Bigl(\frac{r-1}{r}\Bigr)^r\Bigr) - \sum_{q=0}^{r-1}\mathbf F(r,q)P\{q\}\Bigg]  
\end{align}
where
\begin{align}
\mathbf F(r,q) = \Bigl(1-\Bigl(\frac{r-1}{r}\Bigr)^r\Bigr)-\Bigl(1-\Bigl(\frac{r-1}{r}\Bigr)^q\Bigr)  \label{Frq}
\end{align}
The utilization per bank from one interconnect network attached to that bank is then obtained by dividing by $r$.
\begin{align}
\mathbf E_B(n,r) = 1-\Bigl(\frac{r-1}{r}\Bigr)^r -\sum_{q=0}^{r-1} \mathbf F(r,q)P\{q\}  \label{k_port_u_simpler}
\end{align}
In our DSMC we connect $r$ speed-up networks with one from each building block to share and access the $nr$ banks together. This produces an $nr\times nr$ interconnect in a distributed and modular manner. The bank utilization is 
\begin{equation}
\begin{split}
\mathbf U_B(n,r) = 1- \Bigl(\Bigl(\frac{r-1}{r}\Bigr)^r + \sum_{q=0}^{r-1} \mathbf F(r,q)P\{q\}\Bigr)^r  \label{r_speedup_bank_utilization}
\end{split}
\end{equation}
Alternatively, the utilization of a memory bank for a fully connected $nr$ port to $kr$ port topology is \cite{7167206}.
\begin{equation} 
\mathbf {U_{flat}}=1-\bigg(1-\frac{P_a}{kr}\bigg)^{n}\xrightarrow[n=k\to\infty]{} 1-e^{-\frac{P_a}{r}} \\
\xrightarrow[P_a,r=1]{}0.6321 \label{equ-aver-flat-arch}
\end{equation}
\begin{figure}[h]
	\includegraphics[width=1\linewidth]{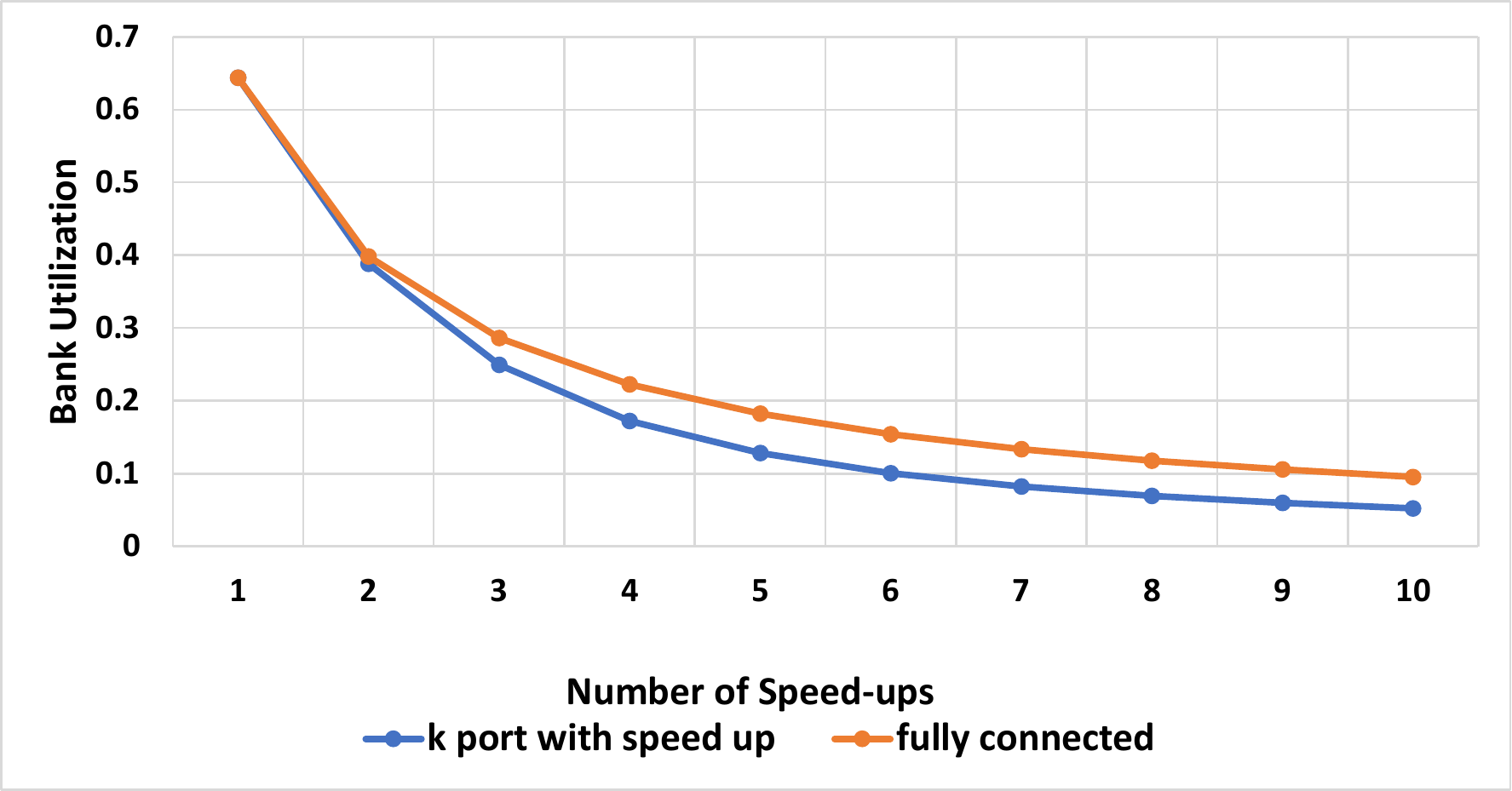}
	\caption{Bank utilization with $n=k=16$ and varying speed up $r$}
	\label{Speedup-BB-Banks} 
\end{figure}

Fig. \ref{Speedup-BB-Banks} shows the case of $n=k=16$ where Formula \eqref{r_speedup_bank_utilization} is the blue line, and Formula \eqref{equ-aver-flat-arch} is the brown line. It shows the utilization of a memory bank in DSMC is lower than that of a fully connected topology. The drop starts from around 1\% per memory bank when $r=2$, to 3.6\% when $r=3$, and 5\% when $r\ge 4$. However, the aggregated utilization per port with speed-up in DSMC is around 77\% when $r=2$, 75\% when $r=3$, 70\% when $r=4$ and 63\% when $r \ge 5$. Correlate this fact with formula \eqref{equ-aver-flat-arch} and wire analyses in section \ref{sec-Related}, we can conclude the cost-effective and beneficial speed-up range for DMSC is from $2$ to $4$ where $r=2$ offers the best cost/performance ratio. Note the above calculations assume random bank access and this is not the optimal memory addressing pattern. To improve it we introduce a \textbf{directed random} scheme in section \ref{sub_Random} to mediate contentions at any memory bank to achieve even higher throughput per port. 

\subsection{Hierarchical and Distributed Architecture with Less Wire Crossings}\label{sub_GI}
Fig. {\ref{simple-BB-16}} shows the basis of a building block that DSMC32M32S is built upon.  Formula \eqref{k_port_u_simpler} and \eqref{r_speedup_bank_utilization} can be applied recursively across stages to calculate the throughput. The reason to use Radix 2 switches is they can significantly simplify the layout by reducing dependency among masters. They are more granular than higher index switches and work easily with irregular geometries on silicon. They also help simplifying memory addressing. Switches with index of 3 and beyond, have $K_{3,3}$ structure as a subgraph \cite{book/218992}, making it very difficult to obtain the isomorphic graphs necessary to ease wire crossings.
For a multi-stage 2-ary interconnect network, we have three types of crossings at each stage as shown Fig. \ref{simple-BB-16}, where each stage has $n/(2g)$ independent blocks, with each block consisting of two independent crossbars that share the next stage input (as in the example above the banks were shared), each with $g$ input and output ports:
\begin{itemize}
	\item Type A: $g/2$ wires going from the masters on left side in a block connect to the slaves on the right side cross the wires from masters on the right connect to the slaves on the left side as shown as the purple dots in Fig. \ref{simple-BB-16}. This happens once per block and leads to $g^2/4$ crossings. 
	\item Type B: A master "self" crossing its own internal master to slave connections in the same side. They are shown as the green dots in Fig. \ref{simple-BB-16}. This happens twice per block and leads to $2*((g/2-1)+(g/2-2)+ \ldots +1) = g(g-2)/4$ crossings in total.
	\item Type C: is a slave "self" crossing where two bundles arriving at the same set of slave ports cross coming from different bundles of master ports. Also shown as the red dots in Fig. \ref{simple-BB-16}. As for type B this leads to $g(g-2)/4$ crossings.
\end{itemize}
The total number of wire crossings among all $n$ masters and $k$ memory ports shown in Fig. {\ref{shared-flat-arch}} is, for $n=k$
\begin{equation}
\mathbf Cr_{n} = \begin{pmatrix} n\\2 \end{pmatrix} \times \begin{pmatrix} n\\2 \end{pmatrix} = \frac{n^2 \times {(n-1)^2}}{4} \backsim O(n^4) \label{equ-radix-N-fully-connected}
\end{equation}
\begin{figure}[h]
	\includegraphics[width=0.8\linewidth,height=0.65\linewidth]{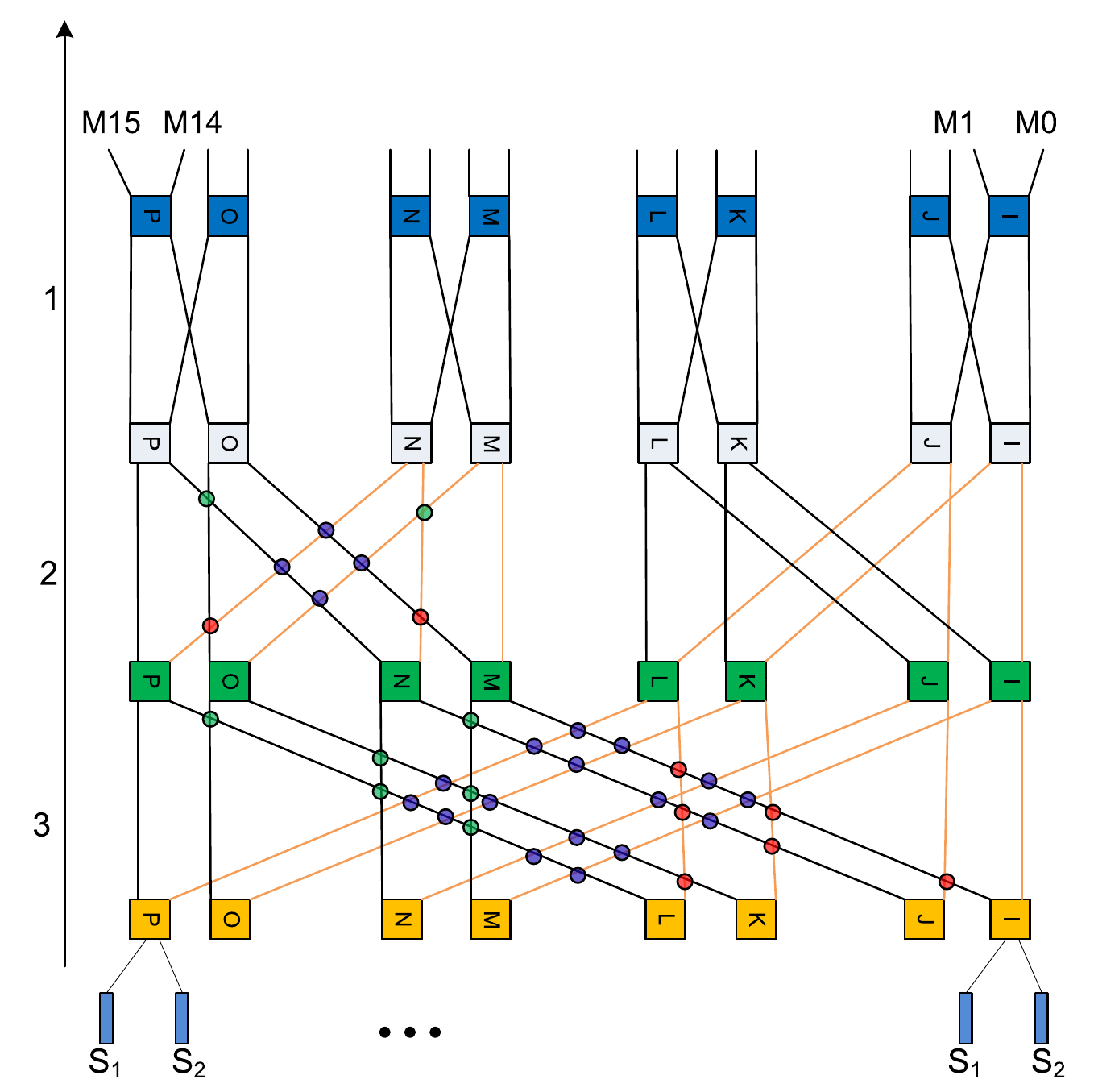}
	\centering
	\caption{A Building Block Based on  A 2-ary 4-fly Network}
	\label{simple-BB-16} 
\end{figure}
Note that in Fig. \ref{simple-BB-16} the number of blocks in each level halves, and therefore $g$ doubles, at each stage. For example, there are $n$ ports in the first stage which has $n/4$ crossbar pairs and $g=n/8$. In general the total crossings of a 2-ary based network is
\begin{align}
\mathbf C_{n}^{(r=2)} = &\sum_{i = 1}^{\log_{2}n - 1} \bigl(\frac{2^i(3\cdot2^i-4)}{4}\bigr)\frac{n}{2^{i+1}} \\
					  = &n\sum_{i = 1}^{\log_{2}n - 1} \bigl(\frac{3\cdot2^{i}-4}{8}\bigr)  \label{equ-radix-2-connected}
\end{align}
Because the "bank sharing" decreases utilization at each stage, we doubled the connections from the second stage onward, so we must multiply the wire crossing count in all stages except the first by 4. This gives the modified crossing count for a building block in DSMC:
\begin{align}
\mathbf C_{n}^{(r=2)} = &n\sum_{i = 1}^{\log_{2}n-1} \bigl(\frac{3\cdot2^{i}-4}{2}\bigr) - \frac{3n}{4}  \label{equ-radix-2-enhanced}
\end{align}
For a $ 32 \times 32$ DSMC, shown in Fig. \ref{Speedup-BB}, we divided it into 2 $16 \times 16$ blocks\footnote{only half of the block diagram is shown because the other half is just the direct mirror image of it} and then connected them 
together with 2 ports going from each stage to its sister stage on the other side, crossing the main highway of $2n$ input ports into the DSMC 4 times and also each other twice. The wire crossings between the two building blocks are:
\begin{align}
\mathbf C_{BxB}^{(r=2)} = 2\bigg[2n+4\sum_{i=1}^{(n/8-1)}(n-8i)\bigg]+\frac{n}{2}  \label{equ-buildingBlock_xrgs}
\end{align}   
Therefore, the total reductions of wire crossing between DSMC with two building blocks sized of $n$ and fully connected cross bar with a size of $2n$ is:
\begin{equation}
\begin{split}
\mathbf R & = \frac{n^2{(2n-1)^2}}{\mathbf 2C_{n}^{(r=2)} +C_{BxB}^{(r=2)}} \\
& = \frac{n{(2n-1)^2}}{\sum_{i = 1}^{\log_{2}n -1} \bigl(3\cdot2^{i}-4\bigr) + 8\sum_{i=1}^{(n/8-1)}(1-8i/n) + 3}  \label{crossing-ratio}
\end{split}
\end{equation}
For this architecture, $n=16$ in formula \eqref{crossing-ratio} gives $R=415.6$. if we reasonably assume that each wire in this calulation represents a bus which normally consists of around 200 physical wires, the physical wire crossing saving is about $400 \times 200^2$, a \textbf{seven} orders of magnitude reduction, leading to significant area saving in the physical implementation results in the next section.
\begin{figure}[h]
	\includegraphics[width=1\linewidth,height=1.15\linewidth]{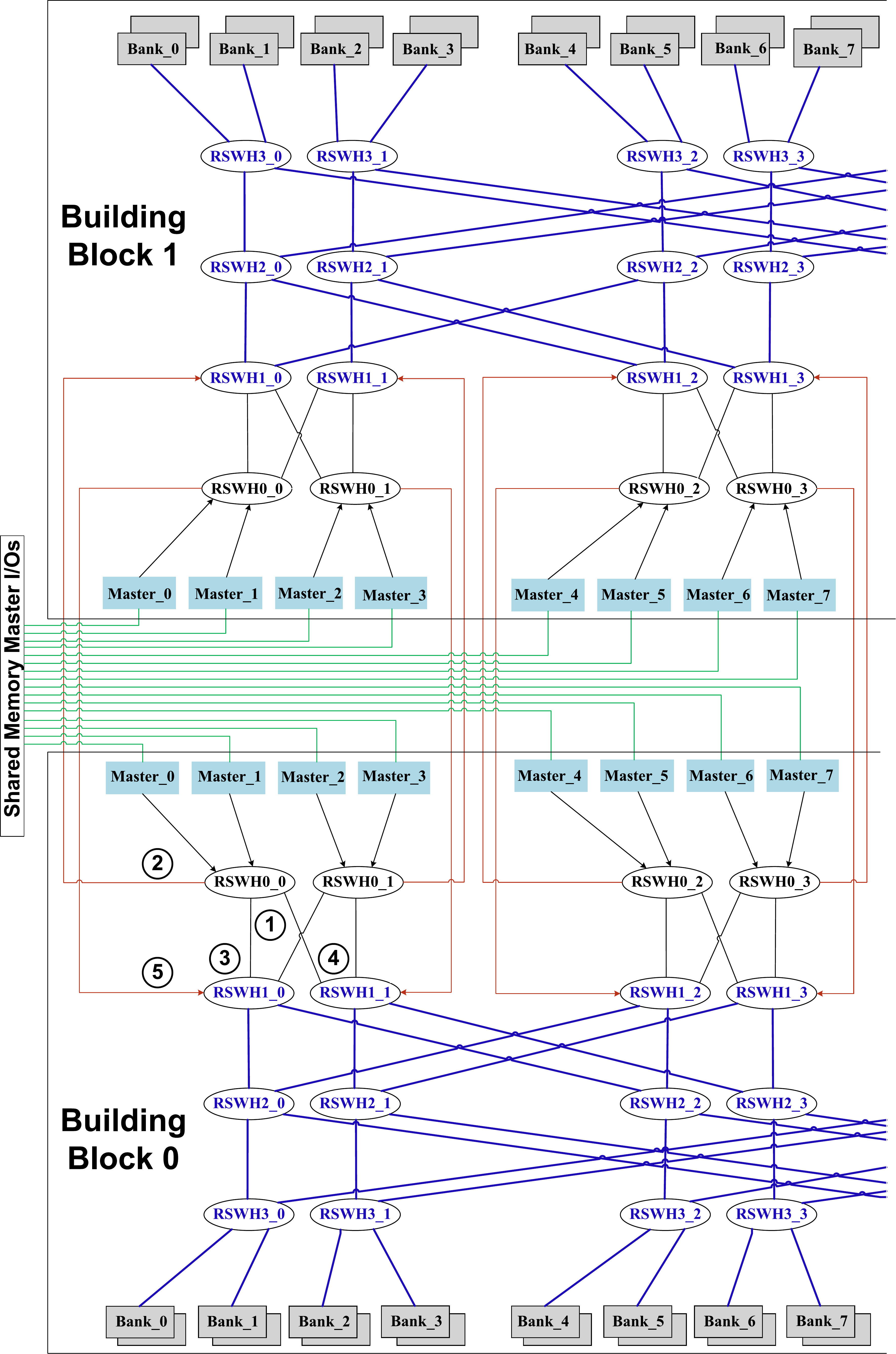}
	\caption{Improve the Overall Throughput by Introducing Speedup Traffic}
	\label{Speedup-BB} 
\end{figure}
\subsection{Whiten and Randomize Traffic with Speed-ups}\label{sub_Random}
 We make up the interconnect throughput loss by adding an extra speed-up network. As shown in Fig. \ref{Speedup-BB} and compared to Fig. \ref{simple-BB-16}, we introduce additional speed-up after the first stage and apply the two level randomization scheme below to keep the port utilization close to 100\% for burst transactions:
\begin{enumerate}
	\item \textbf {Directed randomization at the traffic source}: Disassemble any multi-beat read requests and write data, then spread them evenly across as many building blocks as possible as they enter the shared memory
	\item \textbf{Fractal randomization inside a building block}:  Multiple beats within a linear access are guaranteed to go to different memory banks so to avoid access conflict among themselves 
\end{enumerate}
 Fig. \ref{Speedup-BB} shows how DSMC handles and processes a four-beat request in a configuration with two building blocks. Let{\textquoteright}s assume the request is {\em originally} generated in building block 0, this can be either a read request or a write data transaction and similar requests like this are generated at the same time from other building block as well. The first level radix switch that receives the burst request in building block 0 breaks the even and odd beats of the request into 2 request groups and sends one each to the upper and lower sides (\textcircled{1} and \textcircled{2}, \textbf{directed randomization}). In the lower (upper) side the even (odd) request streams are spread in round robin order to all the blocks of the next stage    for the first two levels of radix switches to make sure two beats within the same burst reach different banks (\textcircled{3} and  \textcircled{4}, \textbf{fractal randomization}) . \\ 
This not only improves average access latency, but will mediate the NUMA effects since it averages out the access latency within a burst request. This is validated by the RTL simulation results in section \ref{results-RTL}.\\
Speed-up is shown in red in Fig. {\ref{Speedup-BB}}. \textbf{All} the first level switches send speed-up traffic to their sister building block as indicated as \textcircled{2} and also receive speed-up traffic as indicated by  \textcircled{5}. All the necessary address decoding for memory banks are done in the first level of the switches in the building block where the requests are originally generated, thus the traffic from other building block(s) can be merged in directly to \textbf{every} the second level switches.

There is an extra independent interconnect network for a speed up of 2 from the level two switches all the way to the memory banks. The blue switches are modified to handle two streams of traffic independently in parallel and the connections among switches and memory banks are all doubled to serve two times of traffic. 
\section{Results}\label{sec-Results}
DSMC-32M32S is prototyped with 4 Mbytes of shared memory and one speedup network ($r=2$) in each building block. Synopsys tool suite is used for synthesis, timing closure and physical design, and the design is timing closed above 600 MHz on a 16nm technology node. The data points for comparisons are  provided by Huawei Wireless in production settings to properly compare our results and show the benefits. We categorize this group of data as Conventional Memory Controller (CMC) data.
\subsection{RTL Simulation Results}\label{results-RTL}
As shown in Fig. {\ref{RTL-read-th}}, the stimulus is generated using uniform random memory access for each traffic pattern\footnote{The mixed traffic has equal percentage of single beat, burst 2/4/8/16 transactions for both read requests and write data.} and the traffic is applied to each and every master port at the same time.
\begin{figure}[h]
	\includegraphics[width=1\linewidth,height=0.5\linewidth]{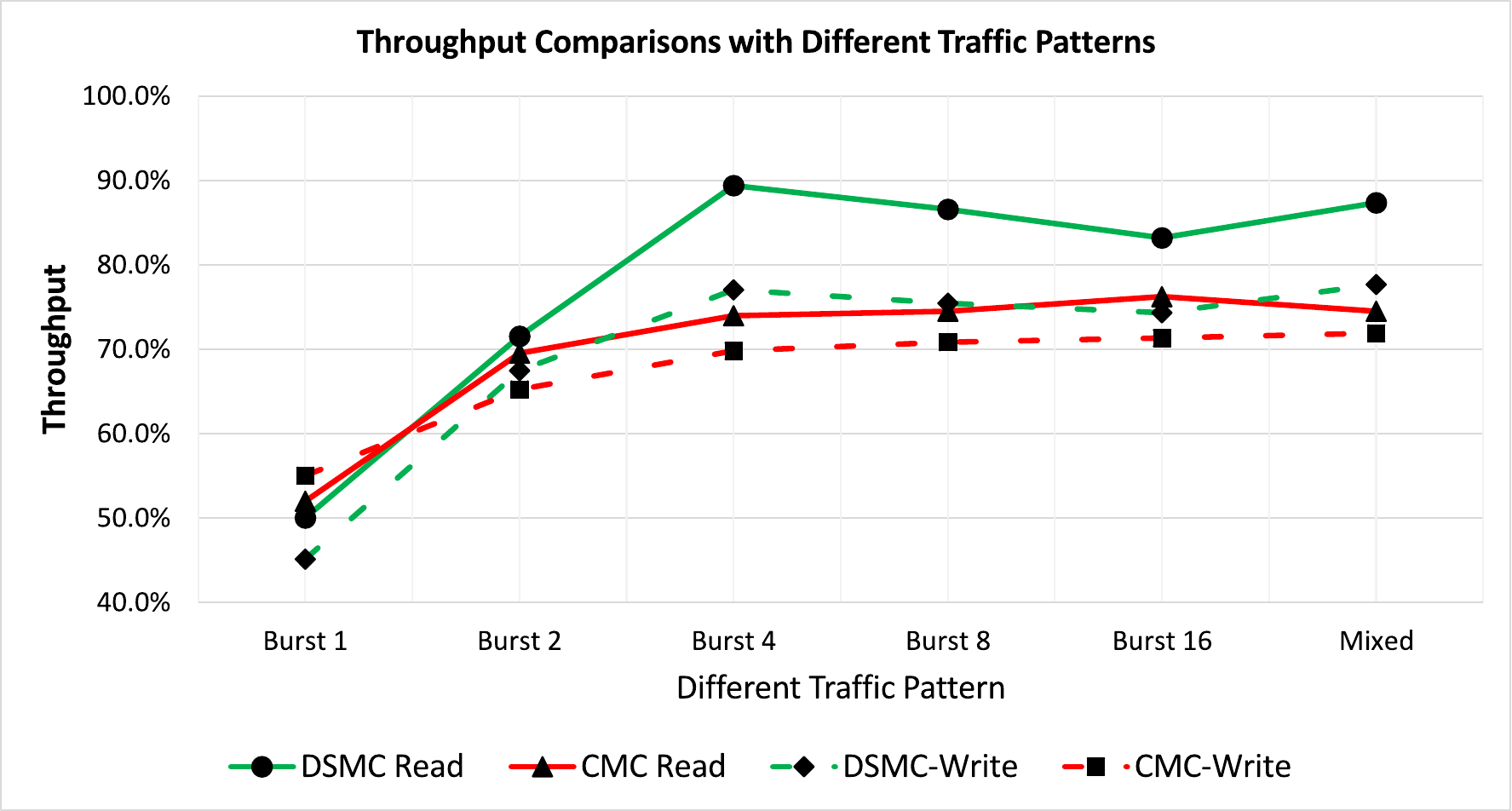}
	\caption{Read and Write Throughput Comparisons for 32M32S Configuration}
	\label{RTL-read-th} 
\end{figure}
Fig. \ref{RTL-read-th} shows the throughput comparisons with different traffic patterns for the read and write accesses. DSMC has almost the same performance when traffic patterns are single and burst 2, but it starts to have over 20\% of combined read and write throughput improvement for the longer bursts beyond 4, and it has about 20\%  improvement for the mixed traffic as well. \\
\begin{figure}[h]
	\includegraphics[width=1\linewidth,height=0.5\linewidth]{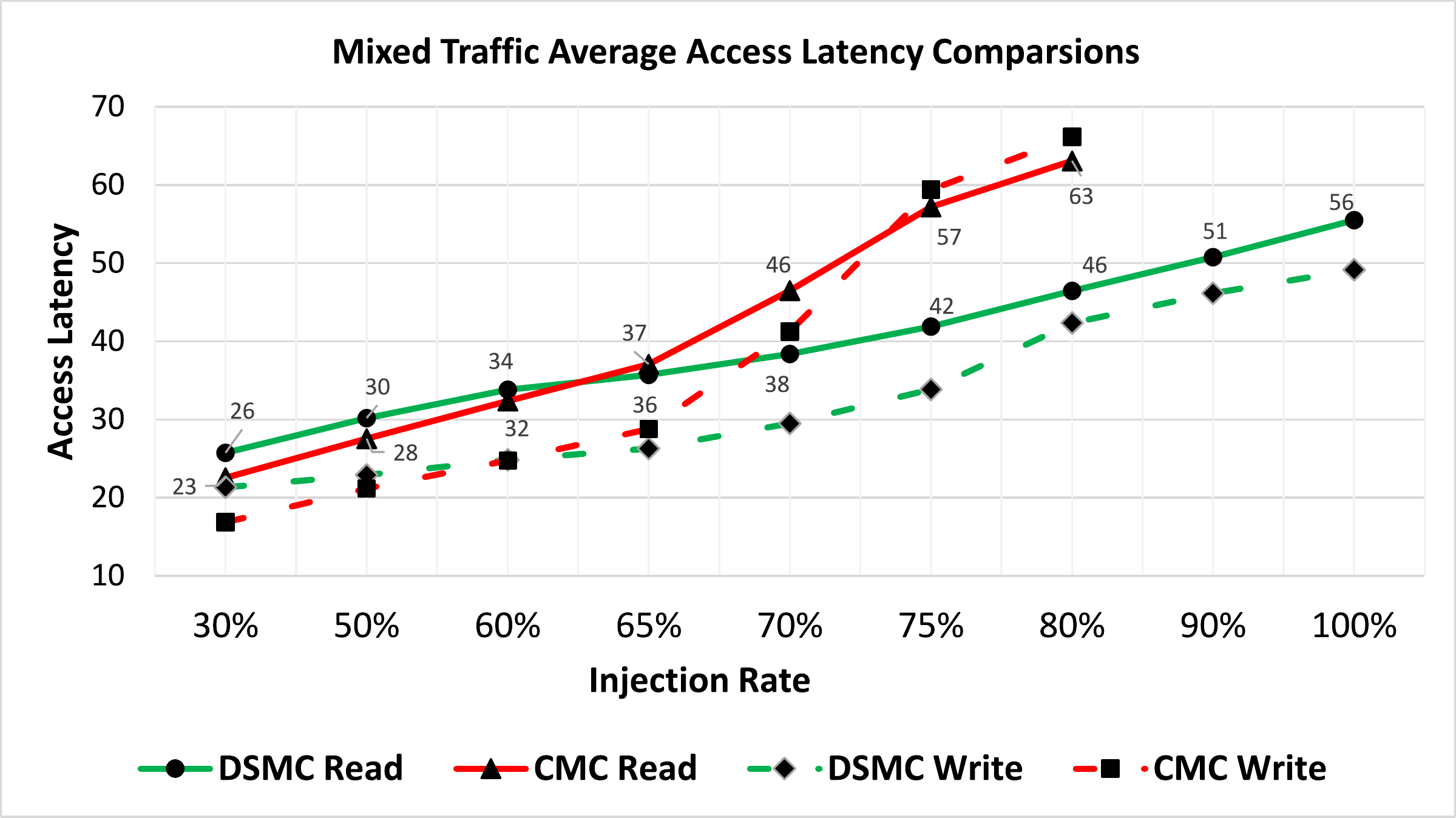}
	\caption{Average Latency Comparisons for 32M32S Configuration}
	\label{RTL-Lat} 
\end{figure}
Fig. \ref{RTL-Lat} shows average latency comparison. As expected, the average latency is almost the same between the two architectures when the traffic load is low. However, the average latency from CMC starts to degrade once the injection rate is over 60\% versus DSMC can handle heavy traffic much better. DSMC has very slow rising curves for both read requests and write data, and the average access latency still maintains less than 60 clock cycles even when 100\% injection rate is applied.\\
\begin{figure}[h]
	\includegraphics[width=1\linewidth]{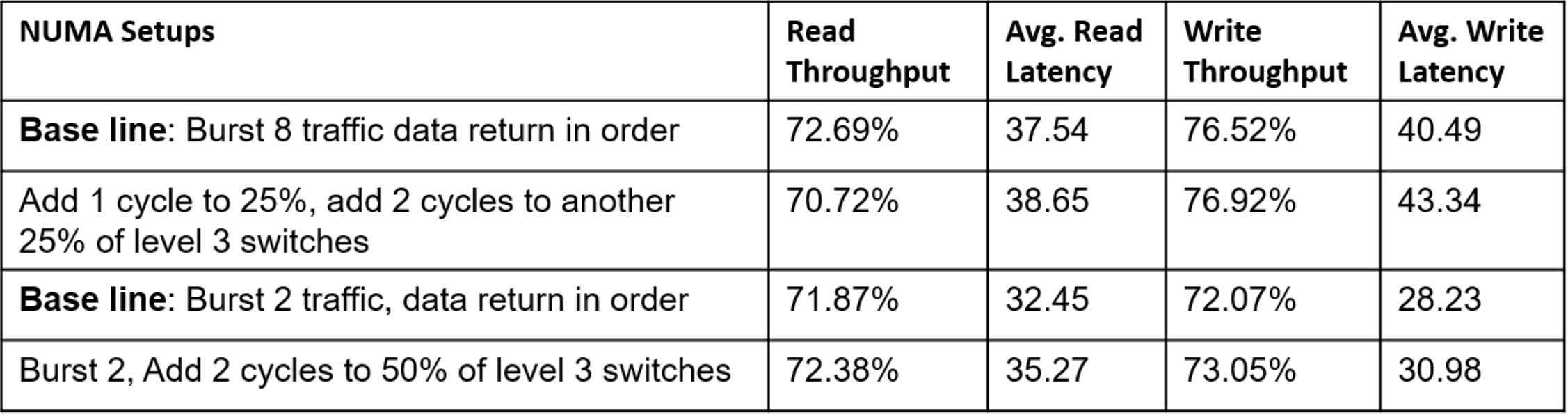}
	\caption{NUMA Effect Mediation after Various Time Closure Slice Insertions}
	\label{RTL-NUMA} 
\end{figure} 
As discussed in section \ref{sec-Related}, sometimes we insert register slices for timing closure due to the widly spread nature of the layout. This makes the design more NUMA. Results are shown in Fig. \ref{RTL-NUMA}. With the randomization techniques described in section \ref{sub_Random}, the write throughput actually improves slightly. The read throughput drops for the burst 8 but burst 2 read improves. This set of results not only correlates well with the findings presented in \cite{Khoretz2013HyperCoreXN}, but also shows that DSMC is resiliant to register slice insertion.
\subsection{Physical Design and Power Estimation}\label{results-PD}

 In a generic industry setting, DSMC-32M32S with two building blocks is always a challenge for any physical design EDA tool. DMSC is very physical design friendly because it works seamlessly with the hierarchical physical design flow. The entire physical design can be split into two parts and be handled in parallel: one for the top level where each building block is abstracted as a black box with accurate I/O timing; the other is at building block level where the majority of design resides. One step further, only \textbf{one} rather than two building blocks needs to go through the physical design. Therefore DSMC can significantly improve end-to-end SoC design experience and overall schedule.\\
Fig. {\ref{DSMC-2-BB-layout}} shows how flexible DSMC can adapt to one of extreme layout constraints where \textbf{all} 32 master access ports are located right in the middle of the left edge. To further stress the architecture, the entire design is taken into the EDA tool all together. A decent result is still obtained because of the physical design awareness of the architecture. The utilization for the area located at middle toward left edge is close to 30\% with less congestion as indicated in the side window. It saves about 30\% silicon area for the {\em interconnect network} compared with that of CMC. The multiple level of simple radix switches enable the data path logic to navigate the sea of memory banks easily to reach all the challenging locations across the layout. This result validates the effectiveness of using lower index switches for timing closure since the timing paths are less dependent on each other. \\
\begin{figure}[h]
	\includegraphics[width=0.7\linewidth,height=0.5\linewidth]{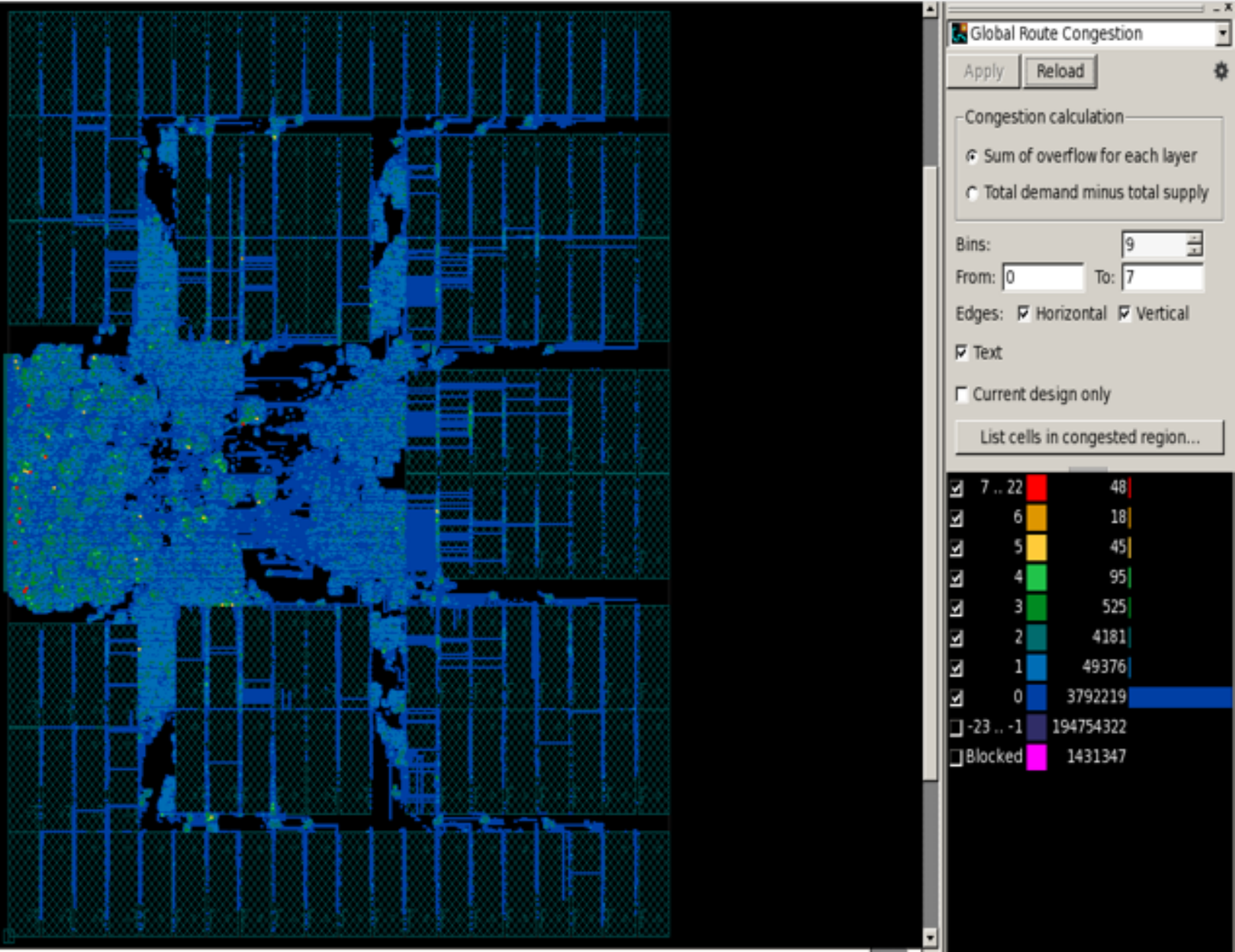}
	\centering
	\caption{The Physical Design View of the DSMC32M32S Configuration }
	\label{DSMC-2-BB-layout} 
\end{figure}
The power consumption is also estimated and compared against the data provided in production setting. The activity vectors are obtained from the same application traces and Synopsys PTPX flow is used to report and compare power consumption. The power consumption is almost the same between DSMC and that of CMC however with different distributions. One observation is that DSMC has 50\% more registers than that of CMC versus CMC has more combinatorial logic than that of DSMC. This result really challenges the industry rule of thumb at least for the wire dominate situation where more registers doesn't always imply more power. 
\section{Conclusions and Future Research Directions}\label{sec-Conclusion}
We present a new shared memory architecture {\textendash} DSMC and validate its benefits. The results show DSMC supports 20\% higher throughput with 20\% lower latency, 30\% less interconnection area and almost the same power than that of the existing  solution. DSMC can easily scale up its capacity by 2x with a much easier implementation flow.\\ 
Moving forward, we actively look forward to further improving the access latency, support even more shared memory and access ports with 3D integration techniques. Meanwhile, we will explore how to leverage this architecture  in the area of Non-Uniform Cache Access because of the similarities between NUMA and NUCA.

\section*{Acknowledgment}

The authors thank Huawei Wireless Division to provide relevant data from production setting to compare and analyze the benefits of this work. The authors thank Xi Chen and Fang Yu who were two major contributors to this architecture exploration; Mr. Xingfeng Chen, Mr. Fa Yin and Mr. Nan Xiang for their outstanding supports in RTL coding and simulations; Dr. Yuchuan Yu, Mr. Bin Yang and Mr. Wei Chen for their valuable inputs on the overall system applications.

\bibliographystyle{IEEEtranS}
\bibliography{refs}

\end{document}